# Spectro-temporal study of atoll source GX 9+9 observed with AstroSat

Sree Bhattacherjee*, Arbind Pradhan, Biplob Sarkar

*Corresponding author

Department of Applied Sciences, Tezpur University, Napaam, Assam 784028, India

Email of corresponding author: app21101@tezu.ac.in

*Abstract*— In this work, we performed a spectro-temporal investigation of the low-mass X-ray binary GX 9+9 using the Large Area X-ray Proportional Counter (LAXPC) and Soft X-ray Telescope (SXT) observation on board AstroSat. The source was detected in the soft state during the observation, which results in a disk dominating energy spectrum within the energy range of 0.7–25.0 keV. We carried out the analysis at different flux levels. In the temporal analysis, LAXPC data in all flux levels showed the presence of noise components, describing broad Lorentzian components. We modeled the energy-dependent temporal properties of the source in order to identify the radiative origin of the observed variability. This source is not a well-studied source; hence we attempt to estimate various source characteristics like inner-disk radius, flux, and inner-disk temperature.

## I. Introduction

GX 9+9 is a neutron star low-mass X-ray binary (NS-LMXB) that was identified during the Galactic center's sounding rocket mission [1, 2]. It is an established atoll source since it traces a 'C'-pattern in the hardness-intensity diagram (HID). Atoll sources primarily have two states: (i) the hard/island state (IS), which is observed in lower luminosities; and (ii) the soft/banana state (BS), which is observed in higher luminosities [3]. BS is further classified into upper and lower banana states. In hard state, the Comptonization component is more dominant, while the accretion disk's thermal emission is more prominent in soft spectra. On durations ranging from days to weeks, the atoll sources show significant spectral changes [4, 5, 6, 7]. The rate of accreting mass in such sources is believed to be increasing from island to upper banana state [3, 8].

The source GX 9+9 has 4.2-hour X-ray and optical waveband modulation [2, 9, 10]. Its distance is asummed to be in between 5 kpc and 10 kpc [10, 11]. The estimated values for the mass and radius of the compact NS in the XRB system are 0.2-0.45 $M_\odot$ and 0.3-0.6$R_\odot$, respectively [2, 9]. It is often detected in a luminous soft state with a ~220 mCrab of X-ray flux in the 2–20 keV band [12]. The source spectrum has been modeled using various spectral descriptions, including a power law [13], a blackbody and Comptonization model [14], a disk component with a Comptonized blackbody [11], and a double blackbody with a broad iron line [15]. Recent studies emphasize the role of spectral modeling in understanding coronal plasma, though these models often have geometric and locational ambiguities [12]. Chatterjee et al. [16] presented Imaging X-ray Polarimetry Explorer (IXPE) data along with AstroSat spectral data from 2020, showing the spectro-polarimetric studies inferring coronal geometry. A simultaneous X-ray and optical band timing investigation for GX 9+9 was conducted using the South African Astronomical Observatory (SAAO) and Rossi Timing Explorer (RXTE) [14]. Over RXTE, LAXPC has a number of benefits, such as, larger effective area, wider energy coverage, etc. Hence, more precise information on the time delays, spectral distribution for a wider energy band, and other timing aspects can be anticipated from AstroSat data.

In this work, we study the specto-temporal properties of the source using AstroSat data and we compare the constancy between the spectral and temporal analysis. We also estimate the characteristics values of the source with the help of the obtained spectral parameters and compare with the previously reported results.

## II. Observation and Data Reduction

AstroSat is India's first multi-wavelength observatory. For this work, we used the Large Area X-ray Proportional Counter (LAXPC) and Soft X-ray Telescope (SXT) instruments of AstroSat. There is a single AstroSat archival observation available for this source (Obs. ID: A09_018T01_9000003774), which was taken on July 25, 2020. The SXT and LAXPC data were utilized for spectral analysis. For timing analysis, we considered only the LAXPC data due to its high timing resolution (in order of microseconds) [17, 18, 19].

LAXPC has three identical counters, of which the LAXPC counters 10 and 30 suffered abnormal gains from 2018 onwards [20]; hence, we only considered the LAXPC 20 data for our analysis. LAXPC operates over a wide energy band from 3-80 keV [17, 18, 19]. We used the South Atlantic Anomaly and Earth occultation-free good time interval data to extract the scientific level-2 data [18], using the standard pipeline of LAXPCsoftware (version: Oct. 13, 2022). The corresponding subroutine of the software has been used to extract the scientific products like light curves, spectrums, background spectrums, and response files.

SXT is an imaging instrument of AstroSat working in the low energy range of 0.3-8.0 keV [21, 22]. The level-1 SXT data for the source has been processed through the standard AS1SXTLevel2-1.4b SXTpipeline to generate the cleaned level-2 event files. The standard background (SkyBkg_comb_EL3p5_Cl_Rd16p0_v01.pha) and response file (sxt_pc_mat_g0to12.rmf) have been used as provided by the SXT payload operational center (POC) to carry out the analysis. During the observation, the SXT count rate exceeded the pileup threshold of 40 counts/sec. In order to limit the counts/sec below the pileup limit, we utilized an annular zone with an outer radius of 13 arcmin and an inner radius of 2 arcmin to extract the light curves and the spectrum utilizing XSELECT (V2.5b). Ancillary response files were created using the task sxtARFModule. We incorporated 2% model systematic error for the background uncertainty estimation [19]. A gain fit has been applied to the SXT data, with a slope of 1 and keeping the offset parameter free, whose value usually lies within 0.02-0.09 [23, 24]. Both the LAXPC and SXT spectrums are optimally grouped using the ftool ftgrouppha.

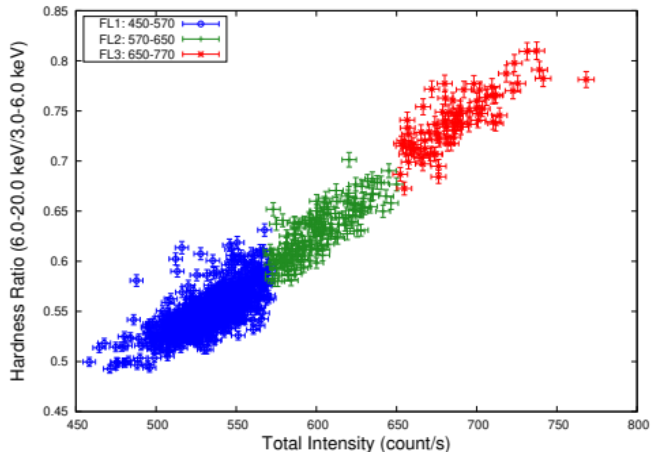

Fig.1. HID of the source using the AstroSat data, with a binning time of 32s. Each color represents the different flux levels as indicated in the plot.

### III. FLUX-RESOLVED SPECTRAL ANALYSIS

This work combines the SXT and LAXPC spectra, utilizing the broad-band spectral coverage from soft to hard X-ray energy range. The HID of the source is displayed in Fig. 1, where the overall count rate in the 3−20 keV band is plotted against the X-ray photon count ratio in the energy ranges of 6–20 keV and 3–6 keV. We may infer that the source was in BS at the time of observation based on the form of HID. As shown in Fig. 1, we segregated the flux of the source into three distinct flux levels (FLs): FL1-450 to 570 counts/s, FL2-570 to 650 counts/s, and FL3-650 to 770 counts/s, in order to conduct the flux-resolved spectral analysis. We have independently undertaken the spectral analysis in each of the FLs within the energy range of 0.7–25 keV in order to investigate the spectral parameter variations with the rise in X-ray flux. For the purpose of the spectral fitting, we use the XSPEC (version: 12.13.0c) [25] package of HEASOFT (6.31.1).

We used the strict simultaneous data from LAXPC and SXT, i.e., when both instruments were observing the source at the same time, for spectrum analysis. First, we consider the blackbody surface (bbodyrad) to be the source of seed photons. However, with a reduced $\chi^2$ value > 2, this scenario was unable to explain the observational data. Next, we proceeded with the assumption that the soft photon was emitting from the accretion disk (diskbb), which gave us a good-fit with physical spectral parameters. Therefore, we obtain const*tbabs*(thcomp*diskbb) as the best-fit model for analysis. To calibrate the different instruments utilized for the joint spectral fitting, we incorporate a cross-calibration component 'const' in the model. We considered the LAXPC 20 spectra as the base spectra, so we kept its constant fixed at 1, leaving the SXT constant free for the calibration. The best-fit model is comparable to the one employed by Chatterjee et al. [16]; however, we have used the thermal Comptonization model (ThComp) [26], an improved version of the nThComp. We extend the energy range with 1000 log bins by "energies 0.01 500 1000 log" for the convolution model ThComp.

Covering fraction, a model parameter of ThComp, indicates the amount of seed photons that will be Comptonized.

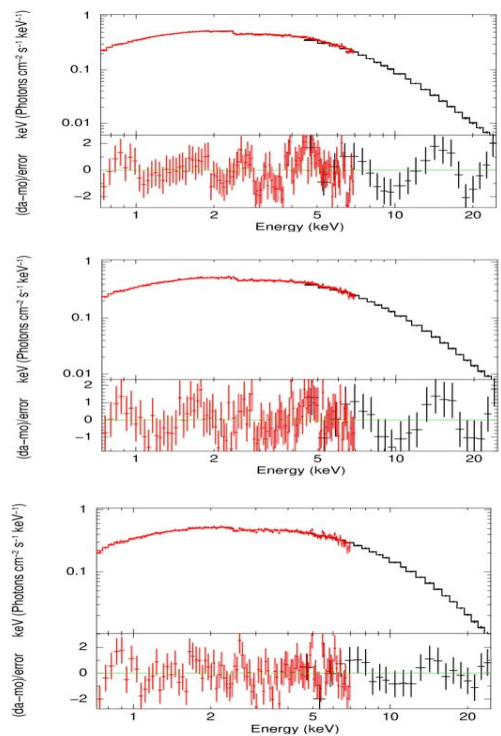

Fig.2. Best-fit spectrum of FL1, FL2, and FL3 from top to bottom, in the energy range 0.7-25 keV. The red and black points represent the SXT and LAXPC 20 data, respectively.

The best-fit spectrum is shown in Fig. 2, while Table 1 displays the best-fit parameters for each FL. The level of confidence for the parametric errors is 90%. To constrain the other parameters, we freeze the source's optical depth at 9. We require edge models (instrumental features) at energy 2.4 keV in order to obtain a better fit [16].

Based on the best-fit parameters from the spectral fitting, we estimated the average inner-disk radius ($R_{in}$) of the source during the observation to be ~ 16.57±1.29 km. It is calculated using the relation,

$$R_{in}= k^2 (N_{dbb}/cos\theta)^{1/2} * D_{10} \quad (1)$$

where, k is the color-correction factor, having a value of 1.7 [27]. $N_{dbb}$ is the disk-normalization, θ is the source's disk-inclination angle, which is considered as 60 degrees [2, 9, 11]. D represents the source distance, which is considered as 10 kpc [11] while $D_{10}$ refers the source distance in the units of 10 kpc. Additionally, using the XSPEC model 'cflux' we estimated the total unabsorbed flux along the FLs in the extended energy range of 0.01-100 keV, which came out to be ~5.94×$10^{-9}$ ergs $s^{-1}$ $cm^{-2}$, 6.51×$10^{-9}$ ergs $s^{-1}$ $cm^{-2}$, 7.30×$10^{-9}$ ergs $s^{-1}$ $cm^{-2}$ in FL1, FL2, and FL3 respectively.

### IV. TIMING ANALYSIS

There are two types of variability that LMXBs generally show: quasi-periodic oscillations (QPOs) and broad-band noise (BBN) features. A power density spectrum (PDS) plot is used to detect the variability of the source during the course of observation. We checked for the variability in different flux levels (FL1, FL2 and FL3) by using the multi-lorentzian approach to the PDS. We didn't detect any hint of QPOs or prominent BBN feature. Further, we decided to look for any statistics as compared to discrete FLs. We generated PDS of full flux in an energy range of 3-20 keV in a frequency range of 0.003 Hz to 1 Hz because of very low statistics beyond 1Hz and below 0.003 Hz. We observed a BBN like feature in a wide centered region as shown in Figure 3 (left). We required two Lorentzians to fit the PDS. We obtained the rms and time lag spectra using LAXPC subroutine "laxpc_find_freqlag" at the centroid frequency ($\nu_o$) as obtained from the fitting which is ~0.04 Hz. The characteristic frequency ($\nu_c$) thus obtained is ~0.1 Hz (see Figure 3 (left)) on applying $\nu_c = \sqrt{(\nu_o^2 + HWHM)}$ [28]. Here, HWHM, is the value indicating the half width of the spectral feature at half its maximum height, is approximately 0.09 Hz. Hence, the peak in the spectra does not align with the $\nu_o$ of the Lorentzian; rather, it is equal to or greater than $\nu_o$. While this difference is minimal for QPOs, which have typically narrow width. But it becomes more pronounced in cases involving BBN features as it spreads over a certain range of frequency [28].

From the Figure 3 (right), we can see that the rms at ~0.1 Hz is increasing as the energy is increasing whereas the phase lag is showing a decreasing trend with energy. But we cannot rely on the phase lag here as the error-bar are consistent within data points. To understand the radiative origin of the observed variability, we used the model as formalized by Garg et al. [29]. It is one of the propagative models based on the assumption of the truncated disk and a hot electron plasma (corona) in between the inner radius and ISCO in a binary system. Where the coronal region is a Comptonizing medium and the soft seed photons originate from the truncated disk. It relates the spectral parameters with the physical parameters, which helps in understanding the factor leading to such stochastic variability in the system.

On the basis of radiative mechanisms, Garg et al. [29] tries to understand the origin of the variability, like QPOs and BBN. It has been successfully incorporated in sources like GRS 1915+105 [29], MAXI 1535-571 [30], H 1743-322 [31] and the 4U 1608-52 [32] to explain the energy variability in these sources.

In order to incorporate this technique, we fit the model as formalized by [29] in rms and phase lag spectra. The relation between the phase lag and the time lag can be given as: time lag is equal to phase lag/$2\pi f$, where f is the peak/central frequency which is usually in millisecond (ms).

We started with the simple scenario where we are changing the inner-disk temperature ($kT_{in}$) and fractional scattering (fs) with some time-delay with respect to the fs.

With this model we were able to obtain a reduced $\chi^2$ value of ~0.80. We further introduced the combination of heating rate, optical depth to check the better fit in rms and lag but we end up getting insignificant $\chi^2$ values (>2). Hence, the best-fit model from this formalism was the variation in $kT_{in}$ and fs with some time delay (phase lag) w.r.t fs, as shown in black solid line in Figure 3 (right). The parameters values of the best-fit are shown in Table 1. As per the analysis, we can infer that the variation responsible for this observed variability was initially introduced as a variation in the inner-disk temperature and the perturbation moved towards coronal region as the variation in the fraction of seed photons that is being Comptonized which can be seen as the positive value of the phase lag. We didn't consider further additional variations of parameter because of bad $\chi^2$ values. So we obtained variability observed in $Kt_{in}$ w.r.t. fs.

TABLE 1. BEST-FIT PARAMETERS OF SPECTRAL ANALYSIS AND ENERGY-DEPENDENT TEMPORAL MODELING

| Spectral Analysis | | | | |
|---|---|---|---|---|
| Par. | FL 1 | FL 2 | FL 3 | Full flux |
| $N_H$ ($10^{22}$ cm$^{-2}$) | 0.16 ± 0.01 | 0.16 ± 0.02 | 0.16 ± 0.02 | 0.15 ± 0.01 |
| $kT_e$ (keV) | 4.14 ± 0.20 | 4.12 ± 0.26 | 3.69 ± 0.22 | 4.17 ± 0.20 |
| fs | 0.14 ± 0.02 | 0.17 ± 0.04 | 0.34 ± 0.10 | 0.15 ± 0.03 |
| $T_{in}$ (keV) | 1.87 ± 0.03 | 2.02 ± 0.06 | 2.04 ± 0.12 | 1.95 ± 0.04 |
| $N_{dbb}$ | 19.67 ± 1.35 | 16.07 ± 1.68 | 15.57 ± 2.03 | 16.96 ± 1.15 |
| $R_{in}$ (km) | 18.18 ± 0.62 | 15.96 ± 1.02 | 15.50 ± 1.92 | 16.83 ± 0.54 |
| $\chi^2$/dof | 152.95/118 | 103.86/111 | 120.96/109 | 130.33/117 |
| Energy-dependent temporal modeling | | | | |
| Varying Par. | $\delta kT_{in}$ (%) | $\delta fs$ (%) | Phase lag | $\chi^2$/dof |
| $Kt_{in}$ w.r.t. $\delta fs$ | 0.56 ± 0.10 | 81 ± 32 | 0.11 ± 0.08 | 6.39/8 |

## V. CONCLUSION

In this work, we have used the AstroSat observation to comprehend the source GX 9+9's timing and spectral characteristics, which are as follows:

1) We perform spectral analysis using the SXT and LAXPC simultaneous data. We determined the source's total unabsorbed flux in the 0.01-100 keV band, observing an increase from FL1 to FL3. The maximum flux reached 7.30 × 10$^{-9}$ ergs s$^{-1}$ cm$^{-2}$, and is comparable to 7.8 × 10$^{-9}$ ergs s$^{-1}$ cm$^{-2}$ reported by [12] in the 0.1–100 keV range.

2) Based on the best-fit spectral parameters, the source's average estimated Rin is around 16.54±1.78 km. Which is relatively consistent to the previously reported estimated value of 27±4 km by Iaria et al [12], where they considered θ as ~43° and a distance of 5 kpc.

3) Based on the variation in spectral parameters with increasing X-ray flux, it is suggested that the source is transitioning towards a softer state. This is indicated by a slight decrease in $kT_e$ and an increase in $T_{in}$ across the FLs.

4) We also conducted the timing analysis using the high timing resolution of LAXPC. We detected BBN-like variability in the PDS centering at ~0.1 Hz, and we further modeled the energy-dependent temporal properties to investigate the radiative origin of the obtained variability.

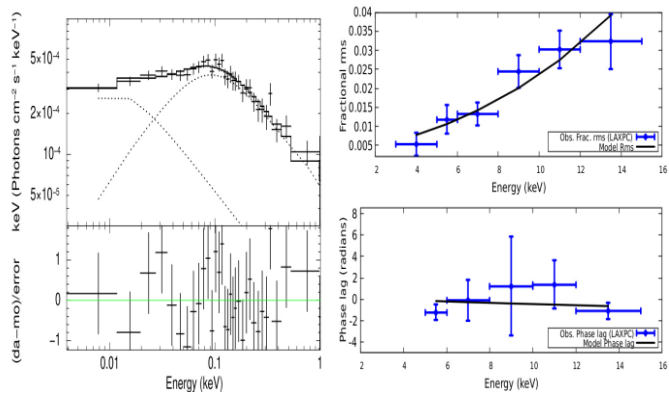

Fig.3. (left): PDS fit and (right): the energy dependent temporal modelling of fractional r.m.s. and phase lag of the representative segment FL1 in top and bottom panel, respectively.

5) Based on the formalism presented by Garg et al. [29], we find that fractional scattering and the change in inner-disk temperature are the causes of the obtained broad-band noise variability.

This work is constrained due to poor data quality of the source. GX 9+9 is not a well-studied source; accordingly, further research into the source is necessary, and a variety of mission observations will be beneficial in order to get a better understanding of the source.

ACKNOWLEDGMENT

This work has used the LAXPC and SXT instruments data of AstroSat mission of ISRO. We thank the LAXPC and the SXT POC at TIFR, Mumbai, for providing the data via the ISSDC data archive and the necessary software tools. Authors acknowledges the academic visits to Inter-University Centre for Astronomy and Astrophysics (IUCAA), Pune for its hospitality and fruitful discussion with the eminent scientists and researcher. B. Sarkar also acknowledge the visiting associateship programme of IUCAA. Author S. Bhattacherjee acknowledge the financial support of DST-INSPIRE (Grant. No.: DST/INSPIRE Fellowship/[IF220164]) fellowship, Ministry of Science and Technology, Government of India. Finally, we would like to acknowledge the anonymous referee for their valuable feedback, which significantly enhanced the quality of the paper.

REFERENCES

[1] Bradt, H., Naranan, S., Rappaport, S., and Spada, G., "Celestial Positions of X-Ray Sources in Sagittarius", *The Astrophysical Journal*, vol. 152, IOP, p. 1005, 1968. doi:10.1086/149613.

[2] Hertz, P. and Wood, K. S., "Discovery of a 4.2 Hour X-Ray Period in GX 9+9", *The Astrophysical Journal*, vol. 331, IOP, p. 764, 1988. doi:10.1086/166597.

[3] Hasinger, G. and van der Klis, M., "Two patterns of correlated X-ray timing and spectral behaviour in low-mass X-ray binaries.", *Astronomy and Astrophysics*, vol. 225, pp. 79–96, 1989.

[4] van der Klis, M., Hasinger, G., Damen, E., Penninx, W., van Paradijs, J., and Lewin, W. H. G., "Correlation of X-Ray Burst Properties with Source State in the ``Atoll'' Source 4U/MXB 1636-53", *The Astrophysical Journal*, vol. 360, IOP, p. L19, 1990. doi:10.1086/185802.

[5] Prins, S. and van der Klis, M., "Correlated X-ray spectral and fast-timing behaviour of 4U 1636-53.", *Astronomy and Astrophysics*, vol. 319, pp. 498–506, 1997. doi:10.48550/arXiv.astro-ph/9701149.

[6] Gierliński, M. and Done, C., "The X-ray spectrum of the atoll source 4U 1608-52", *Monthly Notices of the Royal Astronomical Society* vol. 337, no. 4, OUP, pp. 1373–1380, 2002. doi:10.1046/j.1365-8711.2002.06009.x.

[7] Lin, D., Remillard, R. A., and Homan, J., "Evaluating Spectral Models and the X-Ray States of Neutron Star X-Ray Transients", *The Astrophysical Journal*, vol. 667, no. 2, IOP, pp. 1073–1086, 2007. doi:10.1086/521181.

[8] van der Klis, M., "Similarities in Neutron Star and Black Hole Accretion", *The Astrophysical Journal Supplement Series*, vol. 92, IOP, p. 511, 1994. doi:10.1086/192006.

[9] Schaefer, B. E., "The Optical Light Curve of the Low-Mass X-Ray Binary GX 9+9", *The Astrophysical Journal*, vol. 354, IOP, p. 720, 1990. doi:10.1086/168728.

[10] Christian, D. J. and Swank, J. H., "The Survey of Low-Mass X-Ray Binaries with the Einstein Observatory Solid-State Spectrometer and Monitor Proportional Counter", *The Astrophysical Journal Supplement Series*, vol. 109, no. 1, IOP, pp. 177–224, 1997. doi:10.1086/312970.

[11] Savolainen, P., Hannikainen, D. C., Vilhu, O., Paizis, A., Nevalainen, J., and Hakala, P., "Exploring the spreading layer of GX 9+9 using RXTE and INTEGRAL", *Monthly Notices of the Royal Astronomical Society*, vol. 393, no. 2, OUP, pp. 569–578, 2009. doi:10.1111/j.1365-2966.2008.14201.x.

[12] Iaria, R., "Reflection component in the Bright Atoll Source GX 9+9", *Astronomy and Astrophysics*, vol. 635, 2020. doi:10.1051/0004-6361/202037491.

[13] Church, M. J. and Balucińska-Church, M., "Results of a LMXB survey: Variation in the height of the neutron star blackbody emission region", *Astronomy and Astrophysics*, vol. 369, pp. 915–924, 2001. doi:10.1051/0004-6361:20010150.

[14] Kong, A. K. H., Charles, P. A., Homer, L., Kuulkers, E., and O'Donoghue, D., "Simultaneous X-ray/optical observations of GX9+9 (4U1728-16)", *Monthly Notices of the Royal Astronomical Society*, vol. 368, no. 2, OUP, pp. 781–795, 2006. doi:10.1111/j.1365-2966.2006.10157.x.

[15] Göğüş, E., Alpar, M. A., and Gilfanov, M., "Is the Lack of Pulsations in Low-Mass X-Ray Binaries due to Comptonizing Coronae?", *The Astrophysical Journal*, vol. 659, no. 1, IOP, pp. 580–584, 2007. doi:10.1086/512028.

[16] Chatterjee, R., Agrawal, V. K., Jayasurya, K. M., and Katoch, T., "Spectro-polarimetric view of bright atoll source GX 9+9 using IXPE and AstroSat", *Monthly Notices of the Royal Astronomical Society*, vol. 521, no. 1, OUP, pp. L74–L78, 2023. doi:10.1093/mnrasl/slad026.

[17] Yadav, J. S., "Astrosat/LAXPC Reveals the High-energy Variability of GRS 1915+105 in the X Class", *The Astrophysical Journal*, vol. 833, no. 1, IOP, 2016. doi:10.3847/0004-637X/833/1/27.

[18] Agrawal, P. C., "Large Area X-Ray Proportional Counter (LAXPC) Instrument on AstroSat and Some Preliminary Results from its Performance in the Orbit", *Journal of Astrophysics and Astronomy*, vol. 38, no. 2, Springer, 2017. doi:10.1007/s12036-017-9451-z.

[19] Antia, H. M., "Calibration of the Large Area X-Ray Proportional Counter (LAXPC) Instrument on board AstroSat", *The Astrophysical Journal Supplement Series*, vol. 231, no. 1, IOP, 2017. doi:10.3847/1538-4365/aa7a0e.

[20] Antia, H. M., "Large Area X-ray Proportional Counter (LAXPC) in orbit performance: Calibration, background, analysis software", Journal of Astrophysics and Astronomy, vol. 42, no. 2, Springer, 2021. doi:10.1007/s12036-021-09712-8.

[21] Singh, K. P., "In-orbit performance of SXT aboard AstroSat", in *Space Telescopes and Instrumentation* 2016: Ultraviolet to Gamma Ray, 2016, vol. 9905. doi:10.1117/12.2235309.

[22] Singh, K. P., "Soft X-ray Focusing Telescope Aboard AstroSat: Design, Characteristics and Performance", *Journal of Astrophysics and Astronomy*, vol. 38, no. 2, Springer, 2017. doi:10.1007/s12036-017-9448-7.

[23] Beri, A., "AstroSat observations of the first Galactic ULX pulsar Swift J0243.6+6124", Monthly Notices of the Royal Astronomical Society, vol. 500, no. 1, OUP, pp. 565–575, 2021. doi:10.1093/mnras/staa3254.

[24] Beri, A., "AstroSat and NuSTAR observations of XTE J1739-285 during the 2019-2020 outburst", *Monthly Notices of the Royal Astronomical Society*, vol. 521, no. 4, OUP, pp. 5904–5916, 2023. doi:10.1093/mnras/stad902.

[25] Arnaud, K. A., "XSPEC: The First Ten Years", in *Astronomical Data Analysis Software and Systems* V, 1996, vol. 101, p. 17.

[26] Zdziarski, A. A., Johnson, W. N., and Magdziarz, P., "Broad-band γ-ray and X-ray spectra of NGC 4151 and their implications for physical processes and geometry.", *Monthly Notices of the Royal Astronomical Society*, vol. 283, no. 1, OUP, pp. 193–206, 1996. doi:10.1093/mnras/283.1.193.

[27] Shimura, T. and Takahara, F., "On the Spectral Hardening Factor of the X-Ray Emission from Accretion Disks in Black Hole Candidates", *The Astrophysical Journal*, vol. 445, IOP, p. 780, 1995. doi:10.1086/175740.

[28] Belloni, T., Psaltis, D., and van der Klis, M., "A Unified Description of the Timing Features of Accreting X-Ray Binaries", The Astrophysical Journal, vol. 572, no. 1, IOP, pp. 392–406, 2002. doi:10.1086/340290.

[29] Garg, A., Misra, R., and Sen, S., "Identifying the radiative components responsible for quasi-periodic oscillations of black hole systems", *Monthly Notices of the Royal Astronomical Society*, vol. 498, no. 2, OUP, pp. 2757–2765, 2020. doi:10.1093/mnras/staa2506.

[30] Garg, A., Misra, R., and Sen, S., "On the energy dependence of the QPO phenomenon in the black hole system MAXI J1535-571", *Monthly Notices of the Royal Astronomical Society*, vol. 514, no. 3, OUP, pp. 3285–3293, 2022. doi:10.1093/mnras/stac1490.

[31] Husain, N., Garg, A., Misra, R., and Sen, S., "Investigating the energy-dependent temporal nature of black hole binary system H 1743-322", *Monthly Notices of the Royal Astronomical Society*, vol. 525, no. 3, OUP, pp. 4515–4523, 2023. doi:10.1093/mnras/stad2481.

[32] Bhattacherjee, S., "X-Ray Spectral and Temporal Properties of LMXB 4U 1608-52—Observed with AstroSat and NICER", The Astrophysical Journal, vol. 971, no. 2, Art. no. 154, IOP, 2024. doi:10.3847/1538-4357/ad583d.